\documentclass[12pt,aps,prd,preprint,tightenlines,superscriptaddress,
amsmath,amssymb,nofootinbib]{revtex4}

\RequirePackage[colorlinks=true
,urlcolor=blue
,anchorcolor=blue
,citecolor=blue
,filecolor=blue
,linkcolor=blue
,menucolor=blue
,pagecolor=blue
,linktocpage=true
,pdfproducer=medialab
,pdfa=true
]{hyperref}

\allowdisplaybreaks

\usepackage{amsmath,amssymb,amsthm,amsfonts}
\usepackage{graphicx}
\usepackage{subfigure}
\usepackage{dcolumn}	
\usepackage{hyperref}      	
\usepackage{bm}		
\usepackage{epsfig}
\usepackage{epstopdf}
\usepackage{setspace}
\usepackage[usenames, dvipsnames]{color}
\usepackage{slashed}
\usepackage{comment}
\usepackage{enumitem}
\usepackage[normalem]{ulem}
\usepackage{adjustbox}

\newcommand{\PRE}[1]{{#1}} 

\newcommand{\be}{\begin{equation}\begin{aligned}}
\newcommand{\ee}{\end{aligned}\end{equation}}
\newcommand{\beq}{\begin{equation}}
\newcommand{\eeq}{\end{equation}}
\newcommand{\beqa}{\begin{eqnarray}}
\newcommand{\eeqa}{\end{eqnarray}}

\newcommand{\mev}{\text{MeV}}
\newcommand{\gev}{\text{GeV}}
\newcommand{\tev}{\text{TeV}}

\renewcommand{\eqref}[1]{Eq.~(\ref{#1})}

\newcommand{\order}[1]{\mathcal{O}(#1)}

\def\be{\begin{equation}}
\def\ee{\end{equation}}
\def\bea{\begin{eqnarray}}
\def\eea{\end{eqnarray}}
\def\gsim{\ \rlap{\raise 2pt\hbox{$>$}}{\lower 2pt \hbox{$\sim$}}\ }
\def\lsim{\ \rlap{\raise 2pt\hbox{$<$}}{\lower 2pt \hbox{$\sim$}}\ }
\def\dslash{\kern-4pt \not{\hbox{\kern-2pt $\partial$}}}
\def\pslash{\not{\hbox{\kern-2pt p}}}

\def\gev{{\rm GeV }}
\def\l{{\rm L}}

\definecolor{gray}{rgb}{0.90,1,1}
\definecolor{LightCyan}{rgb}{0.88,1,1}

\def \be{\beta}

\def\beq{\begin{equation}}
\def\eeq{\end{equation}}
\def\bea{\begin{eqnarray}}
\def\eea{\end{eqnarray}}
\def\ber{\begin{eqnarray*}}
\def\eer{\end{eqnarray*}}
\def\bwt{\begin{widetext}}
\def\ewt{\end{widetext}}

\def\roughly#1{\mathrel{\raise.3ex\hbox
{$#1$\kern-.75em\lower1ex\hbox{$\sim$}}}}
\def\lsim{\roughly<}
\def\gsim{\roughly>}

\def\order{\lower 1.8ex \hbox{\LARGE\~{}}}

\usepackage{bm}

\def \({\left(}
\def \){\right)}
\def \[{\left[}
\def \]{\right]}
\def \l|{\left|}
\def \r|{\right|}

\def \be{\beta}

\def\gev{{\rm GeV}}

\def \({\left(}
\def \){\right)}
\def \[{\left[}
\def \]{\right]}
\def \l|{\left|}
\def \r|{\right|}
\def\KLpinunu{ K_L \to \pi^0 \nu \bar{\nu}}
\def\Kpluspinunu{ K^+ \to \pi^+ \nu \bar{\nu}}

\begin{document}

\preprint{UMISS-HEP-2010-02}

\title{\PRE{\vspace*{1.0in}}
Dark sector origin of the KOTO and MiniBooNE anomalies
\PRE{\vspace*{.4in}}}

\author{Alakabha Datta}
\email{datta@phy.olemiss.edu}
\affiliation{Department of Physics and Astronomy, University of Mississippi, 108 Lewis Hall, Oxford, MS 38677 USA
\PRE{\vspace*{.1in}}}
\PRE{\vspace*{.1in}}

\author{Saeed Kamali}
\email{skamali@go.olemiss.edu}
\affiliation{Department of Physics and Astronomy, University of Mississippi, 108 Lewis Hall, Oxford, MS 38677 USA
\PRE{\vspace*{.1in}}}
\PRE{\vspace*{.1in}}

\author{Danny Marfatia}
\email{dmarf8@hawaii.edu}
\affiliation{Department of Physics and Astronomy, University of Hawaii, Honolulu, HI 96822 USA
\PRE{\vspace*{.1in}}}


\begin{abstract}
\PRE{\vspace*{.2in}}
We present a dark sector model that reproduces the KOTO, MiniBooNE and muon anomalous magnetic moment anomalies. The dark sector is comprised of a light scalar singlet $S$ that has a large coupling to a slightly heavier sterile neutrino that mixes with the active neutrinos. The scalar couples to standard model fermions via Yukawa couplings, and to photons via a higher-dimensional coupling.
 The KOTO signal is a result of the 
flavor-changing penguin process $K_L \to \pi^0 S$ followed by the decay of $S$ to neutrinos. 
The sterile neutrino produced  in neutrino-nucleus scattering at MiniBooNE decays to an active neutrino and $S$, which decays electromagnetically and creates an event excess at low energies.

\end{abstract}


\maketitle



\section{Introduction}
\label{sec:introduction}
 Currently, there are many
 measurements in the quark and the lepton sectors that have eluded explanation in the standard model (SM).
 In the quark sector there are several anomalies  in the $B$, $D$ and $K$ systems.  Here we concentrate  on $K$ decays where the interesting modes are the rare kaon decays, $\KLpinunu$ and $\Kpluspinunu$, which are being probed by the  KOTO experiment  at J-PARC and the NA62 experiment at CERN. 
 Recent reports from KOTO~\cite{KOTO1, KOTO2} indicate that $\KLpinunu$ decays occur at  a rate much larger than predicted by the
  SM~\cite{Kitahara:2019lws}. A fair amount of interest has been generated in model building to explain the  KOTO anomaly 
 \cite{Kitahara:2019lws, Egana-Ugrinovic:2019wzj, Dev:2019hho, Li:2019fhz,Jho:2020jsa, Liu:2020qgx,Cline:2020mdt,He:2020jzn,
Ziegler:2020ize, Liao:2020boe, He:2020jly, Gori:2020xvq, Hostert:2020gou}. Based on the number of events observed by the KOTO experiment, the branching ratio can be estimated to be~\cite{Kitahara:2019lws}
\begin{equation}
BR(K_L \to \pi^0 \nu \bar{\nu})_{\mathrm{KOTO}} = 2.1^{+2.0 (+4.1)}_{-1.1 (-1.7)} \times 10^{-9}\,.
\label{eq:KOTOresult}
\end{equation}
This result is two orders of magnitude larger than the SM prediction,
$BR(K_L \to \pi^0 \nu\bar{\nu})_{\mathrm{SM}} = (3.4 \pm 0.6) \times 10^{-11}$~\cite{Buras:2015qea}.

On the other hand NA62 obtains a 90\%~C.L. bound for $\Kpluspinunu$~\cite{NA62},
\begin{equation}
\label{eq:bound_na62}
BR(K^+ \to \pi^+ \nu \bar{\nu})_{\mathrm{NA62}}  <  1.85 \times  10^{-10}\,,
\end{equation}
which appears to violate the Grossman-Nir (GN) bound~\cite{Grossman:1997sk}
\begin{equation}
BR(K_L \to \pi^0 \nu \bar{\nu}) \le 4.3 BR(K^+ \to \pi^+ \nu \bar{\nu})\,. \
\end{equation}
The E787 and E949 experiments at BNL have also measured $BR(K^+ \to \pi^+ \nu \bar{\nu})$~\cite{PhysRevD.77.052003, Artamonov:2009sz}  assuming the pion spectrum predicted by the SM. According to Ref.~\cite{Artamonov:2009sz},
\begin{equation}
BR(K^+ \to \pi^+ \nu \bar{\nu}) = (1.73^{+1.15}_{-1.05}) \times 10^{-10}\,,
\end{equation} 
which is also in conflict with the GN bound.

Many solutions to the KOTO anomaly involve a new light particle $X$ that appears in the decay $ K \to \pi X$, with $X$ decaying outside the detector. As  the KOTO and NA62 detectors have different lengths, by an appropriate choice of parameters,  consistency is achievable. Another option is that if the
$X$ mass is around the pion mass
then there is a range of $m_X$  not probed by NA62  due to the large pion backgrounds from $K^+ \to \pi^0 \pi^+$~\cite{Fuyuto:2014cya}; see  Fig.~2 of Ref.~\cite{Egana-Ugrinovic:2019wzj}.   This gap in sensitivity occurs for $m_X \sim 100-165$~MeV, although
if $m_X$ is very close to the pion mass then part of this gap is covered by a different NA62 analysis, which sets a limit on the invisible decays of the neutral pions from $K^+ \to \pi^+ \pi^{0}$:
$BR[ \pi^0 \to \rm{invisible}] < 4.4 \times 10^{-9}$~\cite{NA62}, which implies $BR[ K^+ \to \pi^+ \rm{invisible}] \sim 10^{-9}$~\cite{Liu:2020qgx}. 
Part of this gap is also covered by E949~\cite{Artamonov:2009sz} which constrains the branching ratio for $K^+ \to \pi^+ X$ as a function of the mass and lifetime of $X$. 

The dark sector model we present in this work has a light scalar, $S$, in the above mass window to avoid the NA62 constraint. 
Because a kinetically-mixed $Z'$ cannot explain the KOTO anomaly~\cite{Jho:2020jsa}, a scalar mediator is an obvious choice.
The scalar interacts with SM particles with coupling strengths proportional to their masses.  Our dark sector also includes a sterile neutrino, 
$\nu_D$, that couples to the scalar with an ${\cal{O}}(1)$ coupling.  The coupling of the scalar to active neutrinos is generated by the mixing of the sterile neutrino with the active neutrinos.
The model generates  the FCNC transitions, $b \to s S$ and $s \to d S$, through the usual penguin diagrams. The corresponding mesonic level process $K_L \to \pi^0 S$ followed by the decay of $S$ to neutrino pairs explains the KOTO measurement.

 The goal of  the MiniBooNE experiment was to address the 3.3$\sigma$ LSND anomaly in electron-like events seen in the $\bar{\nu}_e$ channel~\cite{Aguilar:2001ty}. Over the 15 years of data taken by MiniBooNE, a new anomaly, that is not inconsistent with the LSND anomaly, has gained significance. The data show a 4.8$\sigma$ excess in the low energy part of electron spectra in both the neutrino and antineutrino channels~\cite{Aguilar-Arevalo:2018gpe}. This {\it low-energy excess} begs explanation independently of the LSND anomaly.{\footnote{Accounts of the LSND and MiniBooNE anomalies in terms of oscillations between active neutrinos and an eV-mass sterile neutrino must contend with a raft of experimental constraints, which lead to baroque scenarios as in Ref.~\cite{Liao:2018mbg}.}
Models in which a light neutrino is upscattered into a sterile neutrino which subsequently decays into an $e^+e^-$ pair have been considered in Refs.~\cite{Bertuzzo:2018itn,Ballett:2018ynz} to resolve this anomaly. The mediator through which the light neutrino scatters on the target nucleus is a  $Z'$ boson kinetically mixed with the electromagnetic field tensor.  However, the solution in which the $Z'$ is lighter than the sterile neutrino~\cite{Bertuzzo:2018itn} is excluded~\cite{Arguelles:2018mtc} by data from the CHARM-II~\cite{Vilain:1994qy} and MINERvA~\cite{Valencia:2019mkf} experiments. 
A novel aspect of our model is that electromagnetic decays of the sterile neutrino
produced through neutrino-nucleus scattering via $S$ exchange explains the MiniBooNE anomaly and is compatible with CHARM-II and MINERvA 
data even with $S$ lighter than the sterile neutrino.

There  is  also the long standing anomaly in the
anomalous magnetic moment of the muon, $(g-2)_{\mu}$. 
 The SM prediction~\cite{Blum:2018mom} is $3.7\sigma$ smaller than the experimental measurement~\cite{Bennett:2006fi}:
\begin{equation}
(g-2)_{\mu}^{\text{exp}} - (g-2)_{\mu}^{\text{SM}} = 27.4 \, (2.7) \, (2.6) \, (6.3) \times 10^{-10} \,.
\end{equation}
The first two uncertainties are theoretical and the last, and largest, is experimental.  The experimental uncertainty is expected to be reduced by a factor of four by the Muon $g-2$ Experiment~\cite{Grange:2015fou} at Fermilab, which is currently collecting data.  
With the further addition of a higher dimensional coupling to two photons motivated by recent model building~\cite{Datta:2019bzu} , the 
$(g-2)_\mu$ anomaly can also be addressed in our model. A welcome consequence of this coupling is that the scalar dominantly decays to a photon pair which can be misidentified as electron events and reproduces the MiniBooNE anomaly.

The paper is organized as follows.
 In Section~II, we describe our model and the decays of the scalar and sterile neutrino. In Section~III, we explain the KOTO  and 
 and  $(g-2)_\mu$ anomalies and demonstrate consistency with all relevant constraints. In  Section~IV we  consider  the production of the sterile neutrino in neutrino scattering experiments, and explain the MiniBooNE anomaly.
 We summarize in Section~V.

\section{ Model}
\label{sec:model}

The dark sector has a light singlet scalar $S$, with mass in the range $m_S \sim 100-165 ~\mev$, coupled to a sterile neutrino $\nu_D$ which is heavier than the scalar.
 The scalar has  couplings to SM fermions proportional to their masses:
\begin{equation}
\mathcal{L}_{S} \supset \frac{1}{2}(\partial_\mu S)^2 \! - \! \frac{1}{2}m_{S}^2 S^2 \! - \eta_d\sum_{f=d,l}  \! \frac{m_f}{v} \bar{f}f  S - \eta_u \sum_{f=u} \! \frac{ m_f }{v}\bar{f}f S 
- g_D S \bar {\nu}_D  \nu_D\,.
\label{L_S}
\end{equation}  
Here $v \simeq 246~\gev$, is the vacuum expectation value of the Higgs boson, $d$ and $l$ correspond to down-type quarks and leptons and $u$ corresponds to up-type quarks.   This coupling structure  can arise in ultraviolet complete models in which a light scalar singlet is added to a two-Higgs-doublet model~\cite{Datta:2019bzu, Liu:2020qgx, Batell:2016ove}.
 In this case, the parameters  $\eta_d$ and $\eta_u$ play the role of the mixing parameters between the singlet scalar and the two neutral scalars of the two-Higgs-doublet model.  

The mixing between the flavor eigenstates $\nu_\alpha$ and mass eigenstates $\nu_i$ of the four Dirac neutrinos is given by
\begin{align}
\nu_{\alpha(L,R)} = \sum_{i=1}^{4}U_{\alpha i}^{(L,R)} \nu_{i(L,R)}  ~, \quad (\alpha=e,\mu,\tau,D)\,,
\label{Eq:mixing}
\end{align}   
where $L,R$ denote the handedness of the neutrino, and $U^L$  and $U^R$ are $4\times4$ unitary matrices, which we take to be real and equal ($U^L = U^R\equiv U$). 
Neutrino mixing  induces a coupling of the scalar to light neutrinos.
  $\nu_4$ must be a Dirac neutrino so that its non-relativistic decays, $\nu_4 \to \nu +S$, are not isotropic~\cite{Balantekin:2018ukw}. If $\nu_4$ were Majorana, its decays would be approximately isotropic which is inconsistent with the angular distribution measured by MiniBooNE. 

To address the $(g-2)_\mu$ anomaly we include the higher-dimensional Lagrangian term,
\begin{equation}
\Delta \mathcal{ L_S} =- \frac{1}{4} \kappa S F_{\mu \nu} F^{\mu \nu}\,,
\label{DL_S}
\end{equation}
which yields an $S \gamma \gamma$ coupling governed by the parameter $\kappa$, which has dimensions of inverse mass.  This coupling is generically induced by heavy states, such as leptoquarks, and for $\kappa \sim (1\,\tev)^{-1}$, the light scalar can explain the $(g-2)_\mu$ anomaly~\cite{Datta:2019bzu}. 
In general, a higher dimensional coupling to gluon fields is permitted, which would allow $S$ to decay to hadrons. However, we take $m_S < 2m_\mu$ so that $S$ can only decay to electrons, neutrinos and photons.

The scalar $S$ contributes to $(g-2)_\mu$ via the one-loop and Barr-Zee diagrams~\cite{Barr:1990vd} in Fig.~\ref{fig:g-2}.  The Barr-Zee contribution is induced by the effective $S\gamma\gamma$ coupling which is proportional to $\kappa$.
 The one-loop contribution is given by~\cite{Leveille:1977rc},
\begin{equation}
\delta (g-2)_{\mu}^{(\text{1-loop)}}=\frac{\eta_d^2}{8\pi^2}{m_\mu^2 \over v^2}\int_0^1 dz\frac{(1+z)(1-z)^2}{(1-z)^2+r^{-2} z} \,,
\end{equation}
where $r=m_\mu/m_S$.
The Barr-Zee contribution is dominated by the log-enhanced term~\cite{Davoudiasl:2018fbb},
\begin{equation}
\delta (g-2)_{\mu}^{S\gamma\gamma} \approx \frac{\eta_d}{4 \pi^2} \frac{\kappa m_\mu^2}{v} 
\ln  \frac{\Lambda}{m_S} \,, 
\label{g-2}
\end{equation}
where $\Lambda$ is the cutoff scale which we may take to be of the order of the mass of the particles that induce the effective $S\gamma\gamma$ coupling. We set $\Lambda=2~\tev$. We will see later that the contribution to $(g-2)_\mu$ is dominated by the $S\gamma\gamma$ coupling. 
\begin{figure}
\begin{center}
\includegraphics[width=0.45\textwidth]{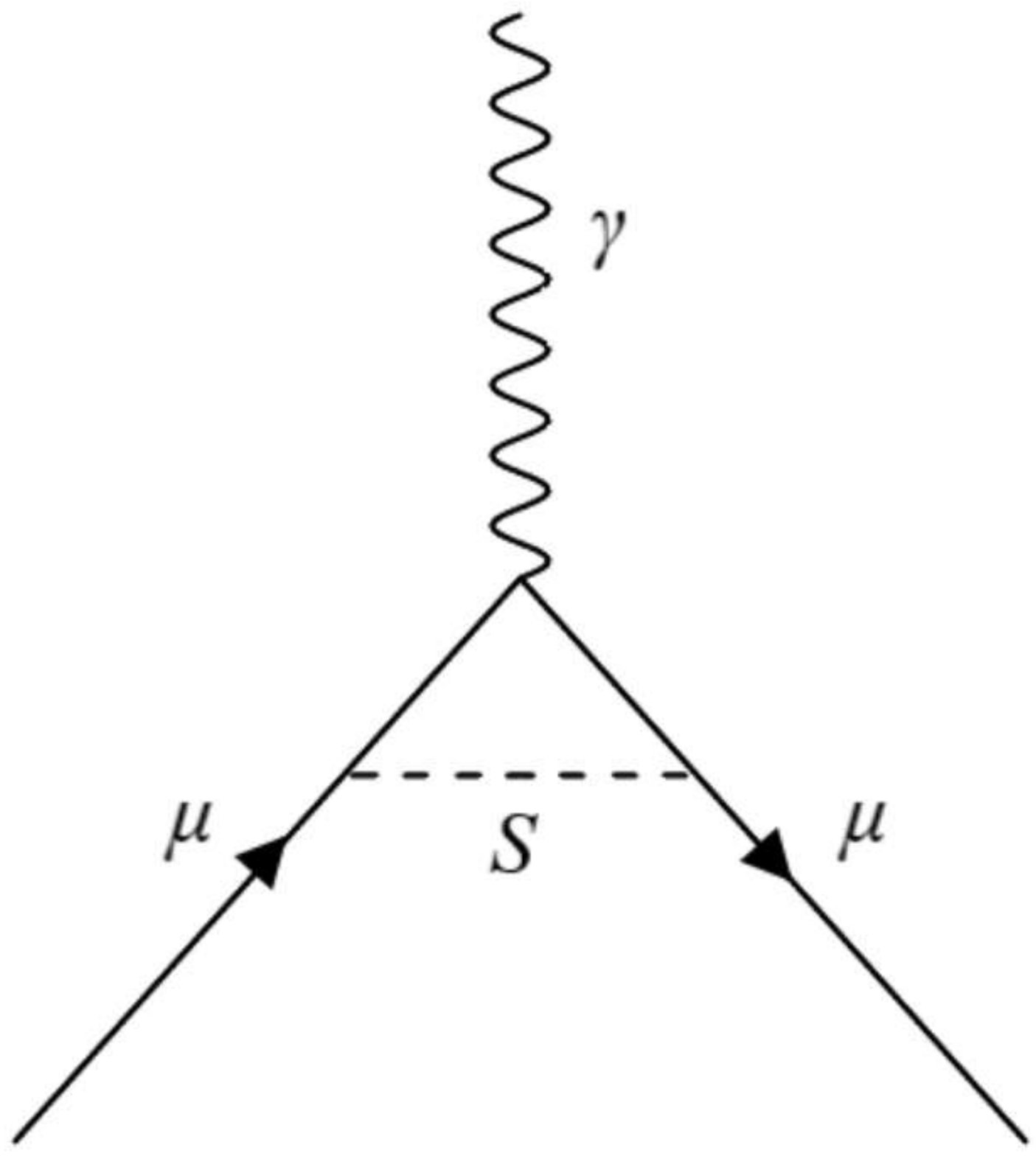}~~~
\includegraphics[width=0.4\textwidth]{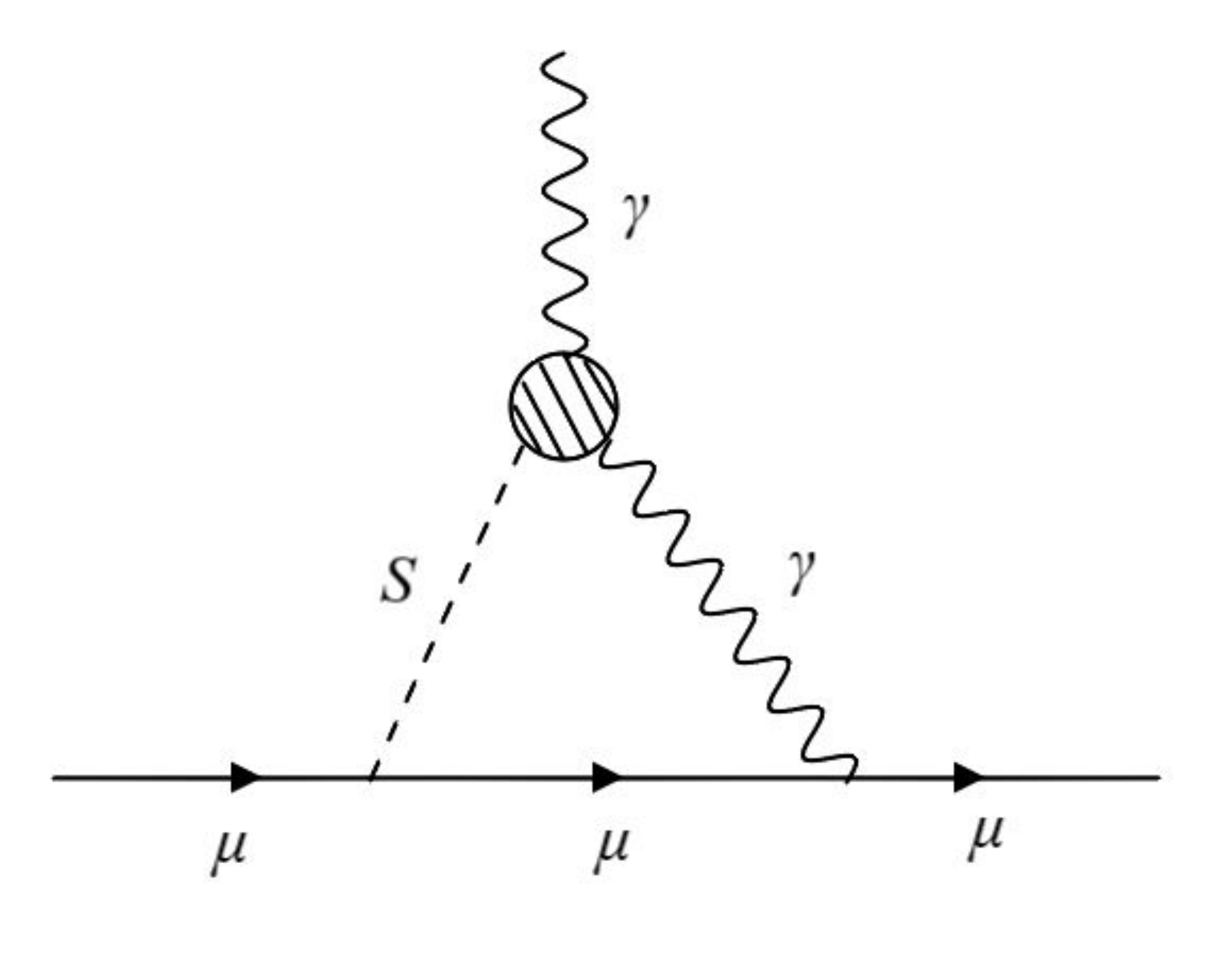}
\caption{The scalar $S$ contributions to $(g-2)_\mu$. }
\label{fig:g-2}
\end{center}
\end{figure}

 The decay  width of $S$ to all three light neutrinos ($\nu_i, ~i=1,2,3$) is 
\begin{align}
\Gamma_{S\to\nu\nu} = \frac{g_D^2}{8\pi}(1-|U_{D4}|^2)^2m_S\,,
\end{align}
and its decay width to $e^+e^-$ is given by
\begin{align}
\Gamma_{S\to e^+e^-} = \frac{\eta_d^2}{8\pi} \frac{m_e^2 m_S}{v^2} \left( 1-4{m_e^2\over m_S^2} \right)^{3/2}  .
\end{align}
An expression for its width to photons can be found in Ref.~\cite{Datta:2019bzu}.

The decay width of $\nu_4$ to $S\nu$ (with $\nu$ denoting all three light neutrinos) is
\begin{equation}
\Gamma_{\nu_4 \to S \, \nu} = \frac{g_D^2}{8 \pi}|U_{D4}|^2 \left( 1-|U_{D4}|^2 \right) \left(1-{m_S^2\over m_{\nu_4}^2}  \right)^2 m_{\nu_4}\,.
\end{equation}
Assuming $U_{e4} \approx U_{\tau4} \approx 0$, by unitarity we have $1-|U_{D4}|^2=|U_{\mu4}|^2$.  Note that the scalar has a much longer lifetime $\sim |U_{\mu4}|^{-4}$ than the sterile neutrino $\sim |U_{\mu4}|^{-2}$. 

\section{KOTO and \boldmath{$(\lowercase{g}-2)_\mu$} anomalies}

The coupling of $S$ to up-type quarks leads to the flavor changing neutral transitions $b \to s$ and $s \to d$ via the penguin loop, thus contributing to several rare hadronic decays.  We examine two cases: 
\begin{enumerate}

\item{$\kappa \neq0$: We consider the full Lagrangian and find the parameter values that can explain the KOTO, MiniBooNE and $(g-2)_\mu$ anomalies. }

\item{ $\kappa=0$: We neglect the effective $S\gamma\gamma$ coupling and find the parameters that can explain the KOTO and MiniBooNE anomalies, but not the $(g-2)_\mu$ anomaly.}

\end{enumerate}

In the $B$ and $K$ systems, the off-shell effects of the scalar mediator are sub-dominant and place only weak constraints on the parameters of the model.  The full list of such constraints is  provided in Ref.~\cite{Datta:2019bzu}. The main constraints therefore come from on-shell production of the scalar $S$. 
 (Obviously, when the $S\gamma\gamma$ coupling is neglected, the constraints with $\gamma \gamma$ final states are not taken into account.) The primary constraints are

\begin{itemize}

\item{$K_L \to \pi^0 e^+ e^-$: We require $BR(K_L \to \pi^0 e^+ e^-) < 2.8 \times 10^{-10}$~\cite{AlaviHarati:2003mr}.}

\item{$K_{L,S} \to \pi^0 \gamma \gamma$: For these decay modes, we take the scalar contribution to be smaller than their measured central values: $BR(K_{L} \to \pi^0 \gamma \gamma) = (1.273 \pm 0.033) \times 10^{-6}$ and $BR(K_{S} \to \pi^0 \gamma \gamma) = (4.9 \pm 1.8 )\times 10^{-8}$~\cite{Tanabashi:2018oca} }.

\item{$K^+ \to \pi^+ \gamma \gamma$: We require the branching ratio to be smaller than the central value of the measurement, $BR(K^+ \to \pi^+ \gamma \gamma) = (1.01 \pm 0.06) \times 10^{-6}$~\cite{Tanabashi:2018oca}. Note that this is a non-resonant measurement that corresponds to diphoton invariant masses above the range of $S$ masses we consider here. }

\item{$K^+ \to \pi^+ \nu \bar{\nu}$: We require $BR(K^+ \to \pi^+ \nu \bar{\nu}) < 10^{-9}$, obtained by NA62 if the scalar mass is close to the pion mass~\cite{NA62}.}

\item{$B \to K^{(*)}\gamma\gamma$: This decay mode has not been measured. We require the scalar contribution to satisfy $BR(B \to K^{(*)}\gamma\gamma)<10^{-4}$ because for a fraction of the events the two photons could be misidentified as a single photon leading to a signal in $B\to X_s\gamma$~\cite{Tanabashi:2018oca}.}

\item{$B \to K^* e^+ e^-$: This decay is measured at LHCb \cite{Aaij:2013hha}. We require the branching ratio to lie within $1\sigma$ of the measured value, $BR(B \to K^* e^+ e^-) = (3.1_{-0.8-0.3}^{+0.9+0.2} \pm 0.2 \pm 0.5) \times 10^{-7}$; the last uncertainty is the theoretical uncertainty.  }

\end{itemize}     

With the full Lagrangian, we resolve the $(g-2)_\mu$ anomaly within $1\sigma$ and the KOTO anomaly at $95\%$~C.L.  
In Table~\ref{table:params_with_kappa} we provide benchmark points that solve the KOTO, MiniBooNE   and $(g-2)_\mu$ anomalies and satisfy the above constraints. Their corresponding branching fractions to various modes are  as in Table~\ref{table:obs_with_kappa}. Note that the 
90\%~C.L. experimental constraint, $B\to K^{(*)} \nu \bar{\nu} < 2.6(1.8) \times 10^{-5}$~\cite{Tanabashi:2018oca}, is easily satisfied by the benchmark points. 
The interesting  signals of the model are $B \to K^{(*)} \gamma \gamma$  and $ K \to \pi \gamma \gamma$ decays via resonant production of $S$, with branching ratios, $\sim 10^{-5}$ and $10^{-7}$, respectively.

\begin{table}[]
\begin{adjustbox}{width=1\textwidth}
\begin{tabular}{|c|c|c|c|c|c|c|c|}
\hline
Benchmark point  & $\kappa(\text{TeV}^{-1})$  &  $\eta_u\times10^{2}$ & ${\eta_d}$ &  $g_D$ & ${U_{\mu 4}}\times10^{3}$ & $m_S(\text{MeV})$   & $m_{\nu_4}(\text{MeV})$  \\
   \hline
1 & 0.42    &  0.20  & 0.56     &  0.29  & 4.5 &  133  & 416  \\
 \hline
2 & 0.87    &  0.75  & 0.27     &  1.9   & 1.2 &  113  & 417 \\
 \hline
3 &  1.4   &  0.22   & 0.15     &  1.2   & 3.9 &  116  & 443 \\
 \hline
4 & 0.61   &  0.31   &  0.39       &  0.59   & 3.0 & 109  & 462  \\
 \hline 
5 &  0      & 0.065  & 0.89     &  3.0   & 2.6 &  134  &   402 \\
\hline
6 &  0      &  0.070 & 0.87     &  3.2   & 2.5  & 129  &  408  \\
\hline
\end{tabular}
\end{adjustbox}
\caption{Benchmark points with $\kappa\neq 0$ solve the KOTO, MiniBooNE  and $(g-2)_\mu$ anomalies and satisfy constraints from kaon decays, $B$ decays, and neutrino scattering data from CHARM-II. For $\kappa=0$, the $(g-2)_\mu$ anomaly is unsolved. }
\label{table:params_with_kappa}
\end{table}

\begin{table}[]
\begin{adjustbox}{width=1\textwidth}
\begin{tabular}{|c|c|c|c|c|c|c|}
\hline
\shortstack{Benchmark point\\$ $ } & \shortstack{$BR(S\to\gamma\gamma)$\\$ $ }  &  \shortstack{$BR(S\to e^+e^-)$\\$\times 10^{3}$}  &\shortstack{$BR(S\to\nu\bar{\nu})$\\$\times 10^{2}$}     &\shortstack{$BR(B\to K^{(*)} S)$\\$\times 10^{5}$}   & \shortstack{$BR(K^+\to \pi^+ S)$\\$\times 10^{7}$} & \shortstack{$BR(K_L\to \pi^0 S)$\\$\times 10^{7}$}  \\
\hline
1  & 0.911 &  3.2  &  8.5  & 0.19 & 0.063 & 0.27  \\
 \hline
2  & 0.994 &  0.26  &  0.54    & 2.7  &  0.92  & 4.0 \\
 \hline
3 &  0.901 &  0.026  &  9.9    & 0.23  &  0.076  &  0.32 \\
 \hline 
4 & 0.946  &  1.2  & 5.2   &  0.45    &  0.15 &  0.65 \\
 \hline 
 \shortstack{\\$ $ } & \shortstack{$BR(S\to\gamma\gamma)$\\$ \times 10^{4} $ }  &  \shortstack{$BR(S\to e^+e^-)$\\$\times 10^{3}$}  &\shortstack{$BR(S\to\nu\bar{\nu})$\\$ $}     &\shortstack{$BR(B\to K^{(*)} S)$\\$\times 10^{7}$}   & \shortstack{$BR(K^+\to \pi^+ S)$\\$\times 10^{10}$} & \shortstack{$BR(K_L\to \pi^0 S)$\\$\times 10^{9}$}  \\
\hline
5 & 5.5  &  8.8  &  0.991  &  2.0  &   6.7  &  2.9 \\
\hline
6 & 4.3  &  7.5  &  0.992  &  2.4  &  7.9  &  3.4  \\
\hline
\end{tabular}
\end{adjustbox}
\caption{Observables for the benchmark points in Table~\ref{table:params_with_kappa}. }
\label{table:obs_with_kappa}
\end{table}

\section{MiniBooNE anomaly}

 The sterile neutrino is produced via  coherent or incoherent scattering of an active neutrino on a nucleus through scalar exchange, $\nu_\mu + N \to \nu_4 + N$. The
effective coupling generated by the interaction term in Eq.~(\ref{L_S}) and neutrino mixing is $g_D U_{\mu4}|U_{D4}|^2 S \, \bar{\nu}_{4R}  \nu_{\mu L}$.  To calculate the coherent scattering cross section mediated by the scalar $S$, we define the coupling between the nucleus and the scalar $C_N$:
\begin{equation}
\mathcal{L}_{SN} = C_N S \bar{\psi}_N \psi_N\,,
\end{equation} 
where $\psi_N$ is the spinor of the nucleus and $C_N$ is related to the couplings of the scalar to the proton ($C_p$) and neutron ($C_n$), 
\begin{equation}
C_N = Z C_p + (A-Z) C_n\,,
\end{equation}
where $Z$ and $A-Z$  are the numbers of protons and neutrons in the nucleus, respectively.  The nucleon couplings are in turn related to the quark-scalar couplings, $\eta_u \frac{m_u}{v}$ for up-type quarks and  $\eta_d \frac{m_d}{v}$ for down-type quarks:
\begin{equation}
C_p = {m_p\over v} \left( \sum_u \eta_u f^p_u +\sum_d \eta_d f^p_d \right)\,, \quad C_n = {m_n\over v} \left( \sum_u \eta_u f^n_u + \sum_d \eta_d f^n_d \right)\,.
\end{equation} 
Here $m_p$ and $m_n$ are the proton and neutron masses, and $f^p$ and $f^n$ are the proton and neutron form factors~\cite{Crivellin:2013ipa,Hoferichter:2015dsa,Junnarkar:2013ac}. Note that for our choice of quark couplings, the nucleon couplings of the scalar are independent of the quark masses.  \\

The coherent scattering cross section is 
\begin{align}
\frac{d\sigma_S}{dT} = \frac{   g_D^2}{16\pi} |U_{\mu4}C_N|^2 |U_{D4}|^4\frac{(2M + T)(m_{\nu_4}^2 + 2MT)}{E_{\nu_\mu}^2(m_S^2+2MT)^2}F^2(T) \,,
\end{align}
where $T$ is the recoil energy, $E_{\nu_\mu}$ is the muon neutrino energy, $M$ is the mass of the nucleus, and  $F(T)$ is the nuclear form factor~\cite{Helm:1956zz}. 
For our benchmark points, coherent scattering dominates incoherent scattering at CHARM-II overwhelmingly, and by about 25\%-50\% at MiniBooNE. We therefore include an incoherent contribution only for MiniBooNE. 

Rather than analyzing MiniBooNE and CHARM-II data, we apply the results of Refs.~\cite{Bertuzzo:2018itn} and~\cite{Arguelles:2018mtc}, which were obtained in the context of a dark $Z'$ mediator kinetically mixed with the electromagnetic field, to our scalar mediator, with suitable modifications. 
To ensure that our model explains the  MiniBooNE anomaly we impose the following constraints:
\begin{enumerate}[label=(\roman*)]
\item{ We require 
$\int \Phi \frac{d\sigma_S}{dT}  dT dE_{\nu_\mu} \times (BR[ S \to e^+ e^-] + BR[S \to \gamma \gamma])$ to be within 5\% of the value of
$\int \Phi \frac{d\sigma_{Z'}}{dT}  dT dE_{\nu_\mu} \times BR[ Z' \to e^+ e^- ]$ found for the $Z'$ benchmark point in Ref.~\cite{Bertuzzo:2018itn} to explain the MiniBooNE anomaly. Here, $\Phi$ is the $\nu_\mu$ flux at the Booster Neutrino Beam in the neutrino run~\cite{AguilarArevalo:2008yp}, and $\sigma_S$ and $ \sigma_{Z'}$ are  scattering cross sections, including coherent and incoherent contributions, for the scalar and $Z'$ mediators, respectively. }

\item{We implement the CHARM-II constraint in Ref.~\cite{Arguelles:2018mtc} (which excludes the $Z'$ model of Ref.~\cite{Bertuzzo:2018itn}) by requiring $\sigma_S \times  (BR[ S \to e^+ e^-] + BR[S \to \gamma \gamma]) <  \sigma_{Z'} \times BR[ Z' \to e^+ e^- ]$ at CHARM-II for $\langle E_{\nu_\mu} \rangle=20$~GeV~\cite{Tanabashi:2018oca}, where the right-hand-side is evaluated for the parameter values in Fig.~3 of Ref.~\cite{Arguelles:2018mtc} with 
$|U_{\mu4}|=10^{-4}$.
}

\item{We require $m_{\nu4}>400$~MeV so that less than 70\% of the excess events are in the most forward bin ($0.8<\cos\theta<1$) of the angular distribution of electron-like events at MiniBooNE~\cite{Arguelles:2018mtc}.}
\end{enumerate}

The benchmark points in Table~\ref{table:params_with_kappa} satisfy these constraints. 
For $\kappa=0$, solutions occur only in narrow parameter ranges. 
A nonzero $\kappa$ opens up the parameter space by facilitating a substantial branching fraction to $\gamma\gamma$ which mimics the MiniBooNE signal. 
Both the scalar and the sterile neutrino are short lived and have rest-frame decay lengths shorter than 0.1~mm, thereby evading  bounds from experiments that probe long lived particles.
While we conclude that our benchmark points resolve the KOTO, MiniBooNE and $(g-2)_\mu$ anomalies, for a full verification a detailed simulation is necessary which is beyond the scope of this work. Solutions that explain the MiniBooNE anomaly and that are compatible with CHARM-II data arise because our mediator is a scalar particle. 
For the light (vector) $Z'$ mediator, the scattering cross section gets enhanced which is in conflict with CHARM-II data for the couplings and mixing needed to explain MiniBooNE data. The difference in cross sections arises from the longitudinal polarization of the $Z'$ propagator
$\sim q^\mu q^\nu/m^2_{Z'}$, where $q$ is the momentum transfer in the scattering process. For the CHARM-II experiment, $M \simeq \langle E_{\nu_\mu} \rangle \simeq 20$~GeV, so  that $\sigma_{Z'}/\sigma_S \sim (M/m_{Z'})^2$ for $m_S \lsim m_{\nu_4} \ll M$.

\section{Summary}
 
 We presented a model with a $100-140$~MeV singlet scalar $S$ and a $400-465$~MeV sterile neutrino $\nu_D$
that resolves the KOTO, MiniBooNE and $(g-2)_\mu$ anomalies.  $S$ couples to $\nu_D$ with an 
${\cal{O}}(1)$ coupling and to  standard model fermions with Yukawa couplings. A higher-dimensional $S\gamma\gamma$ is needed 
to address the discrepancy in $(g-2)_\mu$. 
The scalar couples to active neutrinos through  the mixing of the sterile neutrino with active neutrinos. The model  generates
the FCNC transitions, $b \to s$ and $s \to d$, via penguin diagrams.
 The resulting  $ K_L \to \pi^0 S $ transition  followed by the decay of $S$ to neutrinos  explains the KOTO signal. At MiniBooNE, the sterile neutrino is produced in neutrino-nucleus scattering mediated by the scalar exchange.  The subsequent decay of  the sterile neutrino to an active neutrino and $S$, which in turn decays to  $e^+e^-$ or $\gamma\gamma$, creates
the low-energy excess in the electron-like event data at MiniBooNE. The scenario is compatible with CHARM-II data. 
 Predictions of the model include $B \to K^{(*)} \gamma \gamma$  and $ K \to \pi \gamma \gamma$ decays via resonant production of $S$, with branching ratios, $\sim 10^{-5}$ and $10^{-7}$, respectively.

\bigskip
{\bf Acknowledgments}: We thank Jonathan Feng, Jacky Kumar and Carlos Wagner for useful discussions. 
This work was financially supported in part by
NSF Grant No. PHY-1915142 (A.D.), and DOE Grant No. de-sc0010504 (D.M.).

\newpage
\bibliography{KOTO_new}

\end{document}